\documentclass[11pt]{article}

\usepackage[margin=1in]{geometry}
\usepackage{amsmath,amssymb,amsfonts}
\usepackage{booktabs}
\usepackage{enumitem}
\usepackage{listings}
\usepackage{xcolor}
\usepackage{url}
\usepackage{hyperref}

\lstdefinestyle{rcode}{
  language=R,
  basicstyle=\ttfamily\small,
  breaklines=true,
  columns=fullflexible,
  frame=single,
  numbers=left,
  numberstyle=\tiny,
  keywordstyle=\ttfamily,
  commentstyle=\itshape,
  showstringspaces=false
}

\title{\texttt{regcorr}: An R Package for Regression Models of Pearson Correlation Coefficients}

\author{%
\makebox[0.82\textwidth][s]{%
Ze Lin\hfill Bo Li\hfill Jinyao Shen}%
\\[5pt]
\small School of Statistics, East China Normal University, Shanghai, China
\\[4pt]
}

\date{Version 0.1.0, June 2026}

\begin{document}
\maketitle

\begin{abstract}
Pearson's correlation coefficient is commonly used as a single-number summary of association between two responses. In many applications, however, the strength of association is itself heterogeneous and may vary with demographic, biological, experimental, or environmental covariates. The \texttt{regcorr} package implements regression models in which a Pearson correlation coefficient is linked to a linear predictor of covariates. The package supports bivariate normal responses and bivariate Bernoulli responses, provides Newton--Raphson estimation routines, includes data generators for simulation studies, and supplies a bootstrap-based subroutine for assessing the significance and power of covariate effects. The implementation follows the likelihood-based framework of Dufera, Liu, and Xu (2023) and exposes it through a lightweight R interface with no compiled code and minimal dependencies. This paper describes the statistical model, the computational design of \texttt{regcorr}, reproducible usage examples, and practical guidance for interpreting covariate-dependent correlations. The package is available from the Comprehensive R Archive Network at \url{https://CRAN.R-project.org/package=regcorr} under the MIT license.
\end{abstract}

\noindent\textbf{Keywords:} Pearson correlation; correlation regression; bivariate normal response; bivariate Bernoulli response; Newton--Raphson algorithm; bootstrap inference; R package.

\section{Introduction}

Correlation analysis is one of the first tools used to summarize how two quantities move together. The classical Pearson correlation coefficient is interpretable, scale-free, and bounded between $-1$ and $1$. Its simplicity, however, is also a limitation: in the standard formulation, the correlation between two responses is estimated as a fixed population quantity, even when scientific interest concerns how the association changes across covariate profiles.

Examples arise naturally. In health studies, the association between physical and mental functioning may differ by age, sex, or marital status. In behavioral finance, the association between risk perception and investment behavior may vary across education or income levels. In biological and engineering settings, two measurements may remain marginally stable while their dependence structure changes under different experimental conditions. In these cases, the question is not only whether two responses are associated, but whether the association itself is systematically related to covariates.

Several existing approaches, such as partial correlation and stratified correlation analysis, adjust for covariates or examine separate subgroups. Earlier model-based approaches include Fisher-$z$ transformation models for correlations \cite{bartlett1993} and covariate-adjusted rank-based correlation methods \cite{liu2018}. These approaches are useful but do not directly parameterize the Pearson correlation coefficient as a regression target. Dufera, Liu, and Xu \cite{dufera2023} proposed likelihood-based regression models for covariate-dependent Pearson correlations in two major settings: bivariate normal responses and bivariate binary responses. The R package \texttt{regcorr} turns this methodology into an accessible computational tool.

The contribution of \texttt{regcorr} is software-oriented. It does not propose a new statistical model beyond the cited framework. Instead, it provides a reproducible implementation that lowers the cost of applying, simulating, and teaching regression models of Pearson correlation coefficients. The package is designed around four goals:
\begin{enumerate}[leftmargin=2em]
  \item support both continuous bivariate normal responses and bivariate Bernoulli responses;
  \item provide Newton--Raphson fitting routines for covariate-dependent correlations;
  \item include simulation utilities for generating data under known correlation-regression parameters; and
  \item offer a bootstrap-based simulation subroutine for studying estimation accuracy and testing power.
\end{enumerate}

\section{Statistical background}

Let $(Y_{1i},Y_{2i})$ denote paired responses for subject $i=1,\ldots,n$, and let $x_i=(1,x_{i1},\ldots,x_{ip})^\top$ be a covariate vector including an intercept. The central object is the subject-specific Pearson correlation coefficient
\[
  \rho_i = \mathrm{corr}(Y_{1i},Y_{2i}\mid x_i).
\]
Rather than treating $\rho_i$ as a constant, the model links it to the linear predictor $x_i^\top\beta$ through a monotone transformation
\begin{equation}
  \rho_i = h(x_i^\top\beta),
  \label{eq:link}
\end{equation}
where $\beta=(\beta_0,\beta_1,\ldots,\beta_p)^\top$ governs how covariates affect the association between the two responses.

Two link functions are commonly used in this framework \cite{dufera2023}. If the correlation is assumed positive, a logistic link can map the linear predictor to $(0,1)$:
\begin{equation}
  h_1(t)=\frac{1}{1+\exp(-t)}.
  \label{eq:logistic}
\end{equation}
When both positive and negative correlations are allowed, a Fisher-$z$-type hyperbolic tangent link maps the linear predictor to $(-1,1)$:
\begin{equation}
  h_2(t)=\frac{\exp(t)-1}{\exp(t)+1}=\tanh(t/2).
  \label{eq:tanh}
\end{equation}
In \texttt{regcorr}, these links are selected by the argument \texttt{link = "1"} for the logistic link and \texttt{link = "2"} for the hyperbolic tangent link.

\subsection{Bivariate normal responses}

For bivariate normal responses, suppose that
\[
\begin{pmatrix}Y_{1i}\\Y_{2i}\end{pmatrix}\sim
N\left(
\begin{pmatrix}\mu_{1i}\\\mu_{2i}\end{pmatrix},
\begin{pmatrix}\sigma_1^2 & \rho_i\sigma_1\sigma_2\\
\rho_i\sigma_1\sigma_2 & \sigma_2^2\end{pmatrix}
\right),
\]
where the means and variances are nuisance components and $\rho_i=h(x_i^\top\beta)$ is the parameter of interest. Let
\[
  e_{1i}=\frac{y_{1i}-\mu_{1i}}{\sigma_1},\qquad
  e_{2i}=\frac{y_{2i}-\mu_{2i}}{\sigma_2}.
\]
Ignoring constants not involving $\beta$, the log-likelihood contribution from observation $i$ can be written as
\begin{equation}
  \ell_i(\beta)
  =-\frac{1}{2}\log(1-\rho_i^2)
  -\frac{e_{1i}^2 - 2\rho_i e_{1i}e_{2i}+e_{2i}^2}{2(1-\rho_i^2)}.
  \label{eq:normal_like}
\end{equation}
The likelihood therefore depends on covariates through $\rho_i$, not merely through the marginal means. A positive coefficient $\beta_j$ indicates that increasing $x_{ij}$ increases $h(x_i^\top\beta)$, with the interpretation depending on the selected link function.

\subsection{Bivariate Bernoulli responses}

For binary responses, let $Y_{1i},Y_{2i}\in\{0,1\}$ with marginal probabilities
\[
  p_{1i}=P(Y_{1i}=1),\qquad p_{2i}=P(Y_{2i}=1),
\]
and joint probability $p_{11,i}=P(Y_{1i}=1,Y_{2i}=1)$. The Pearson correlation for two Bernoulli variables is
\begin{equation}
  \rho_i=\frac{p_{11,i}-p_{1i}p_{2i}}
  {\{p_{1i}(1-p_{1i})p_{2i}(1-p_{2i})\}^{1/2}}.
  \label{eq:binary_cor}
\end{equation}
Given $p_{1i}$, $p_{2i}$, and $\rho_i$, the joint cell probabilities can be recovered as
\begin{align*}
  p_{11,i} &= p_{1i}p_{2i}+\rho_i\{p_{1i}(1-p_{1i})p_{2i}(1-p_{2i})\}^{1/2},\\
  p_{10,i} &= p_{1i}-p_{11,i},\\
  p_{01,i} &= p_{2i}-p_{11,i},\\
  p_{00,i} &= 1-p_{11,i}-p_{10,i}-p_{01,i}.
\end{align*}
The binary log-likelihood is then
\begin{equation}
  \ell(\beta)=\sum_{i=1}^{n}\sum_{a=0}^{1}\sum_{b=0}^{1}
  I(Y_{1i}=a,Y_{2i}=b)\log p_{ab,i},
  \label{eq:binary_like}
\end{equation}
where $p_{ab,i}$ depends on $\rho_i=h(x_i^\top\beta)$. In \texttt{regcorr}, correlated bivariate binary data can also be generated using the algorithmic idea of Qaqish \cite{qaqish2003}, exposed through the \texttt{rbinary()} utility.

\section{Estimation and inference}

The package estimates $\beta$ using Newton--Raphson iteration. The large-sample justification follows standard likelihood-based estimation theory \cite{newey1994}. Let
\[
  s(\beta)=\frac{\partial \ell(\beta)}{\partial \beta},\qquad
  H(\beta)=\frac{\partial^2 \ell(\beta)}{\partial\beta\partial\beta^\top}
\]
be the score vector and Hessian matrix, with nuisance parameters replaced by plug-in estimates where needed. The update has the form
\begin{equation}
  \beta^{(m+1)}=\beta^{(m)}-H(\beta^{(m)})^{-1}s(\beta^{(m)}).
  \label{eq:newton}
\end{equation}
The fitted value $\widehat{\rho}_i=h(x_i^\top\widehat{\beta})$ can then be interpreted as the estimated correlation between the two responses at the covariate profile $x_i$.

For hypothesis testing, the relevant scientific questions can be formulated in terms of the components of $\beta$. A global test of covariate effects examines
\[
  H_0: \beta_1=\cdots=\beta_p=0,
\]
whereas an individual covariate test examines $H_0:\beta_j=0$. Because the covariance expression of $\widehat{\beta}$ can be complicated in finite samples, especially in the Bernoulli case, the original methodology recommends bootstrap-based inference for assessing significance. The \texttt{subRoutineTest()} function in \texttt{regcorr} implements a simulation and bootstrap workflow that returns root mean squared error, consistency rate, and empirical power.

\section{Package design}

\texttt{regcorr} is implemented as a lightweight R package for the R statistical computing environment \cite{rcore}. It imports only the base \texttt{stats} package, requires R version 4.1.0 or later, and does not need compilation. The package interface centers on fitting routines, data generators, and supporting utilities. Table~\ref{tab:functions} summarizes the main functions.

\begin{table}[ht]
\centering
\caption{Main functions in \texttt{regcorr}.}
\label{tab:functions}
\begin{tabular}{ll}
\toprule
Function & Purpose \\
\midrule
\texttt{NRfitBivNormal()} & Fit a correlation-regression model for bivariate normal responses. \\
\texttt{NRfitBivBernoulli()} & Fit a correlation-regression model for bivariate Bernoulli responses. \\
\texttt{genDataBN()} & Generate bivariate normal data with covariate-dependent correlation. \\
\texttt{genDataBB()} & Generate bivariate Bernoulli data with covariate-dependent correlation. \\
\texttt{rbinary()} & Generate correlated binary pairs with specified marginal probabilities. \\
\texttt{logistic()} & Compute the logistic transformation. \\
\texttt{subRoutineTest()} & Conduct simulation and bootstrap studies for parameter testing. \\
\bottomrule
\end{tabular}
\end{table}

The two fitting functions return a list containing at least the estimated coefficient vector, the number of Newton--Raphson iterations, and the number of restarts. The design intentionally avoids a large dependency stack. This makes the package easy to install, inspect, and use in teaching or simulation environments. The package also includes safeguards for numerical instability in small-sample bivariate Bernoulli settings, where perfect separation, invalid cell probabilities, or unstable Hessian matrices may otherwise interrupt computation.

\section{Reproducible examples}

\subsection{Bivariate normal responses}

Listing~\ref{lst:normal} shows a minimal workflow. The data generator creates a covariate matrix \texttt{X}, a two-column response matrix \texttt{Y}, and the true covariate-specific correlations \texttt{rho}. The fitting function then estimates the regression coefficients governing the correlation model.

\begin{lstlisting}[style=rcode,caption={Fitting a bivariate normal correlation-regression model.},label={lst:normal}]
library(regcorr)

set.seed(123)
true_beta <- c(0.25, 1, 0)   # intercept and two covariates
true_eta  <- c(0, 0, 0)

dat <- genDataBN(
  numSample = 500,
  p = 2,
  betaTrue = true_beta,
  eta1True = true_eta,
  eta2True = true_eta,
  link = "1"                 # "1" logistic, "2" tanh
)

fit <- NRfitBivNormal(
  Y = dat$Y,
  X = dat$X,
  betaIni = c(0.25, 0, 0),
  link = "1"
)

fit$betaCurrent
fit$numIter
\end{lstlisting}

If the first non-intercept coefficient is positive and large, the model indicates that the corresponding covariate increases the correlation between the two responses. A coefficient near zero suggests little evidence that the covariate changes the association after accounting for the model structure.

\subsection{Bivariate Bernoulli responses}

Listing~\ref{lst:binary} gives the analogous workflow for binary paired responses. The same conceptual model is used, but the likelihood is formed from the four Bernoulli cell probabilities.

\begin{lstlisting}[style=rcode,caption={Fitting a bivariate Bernoulli correlation-regression model.},label={lst:binary}]
library(regcorr)

set.seed(123)

dat_bin <- genDataBB(
  numSample = 300,
  p = 1,
  betaTrue = c(0.3, 0.1),
  eta1True = c(0, 0),
  eta2True = c(0, 0),
  link = "1"
)

fit_bin <- NRfitBivBernoulli(
  Y = dat_bin$Y,
  X = dat_bin$X,
  beta0 = c(0, 0),
  link = "1"
)

fit_bin$betaCurrent
fit_bin$numIter
\end{lstlisting}

Binary-response correlation regression is particularly useful when the outcomes are indicators, such as the presence or absence of two conditions, two behaviors, or two events. In such cases, the model allows researchers to ask whether the dependence between the two indicators varies with subject-level covariates.

\subsection{Simulation-based assessment}

The function \texttt{subRoutineTest()} provides a compact route to simulation-based evaluation. Listing~\ref{lst:simulation} illustrates the syntax for a small demonstration run. In a full simulation study, \texttt{numSimu} and \texttt{numBoot} should be increased substantially.

\begin{lstlisting}[style=rcode,caption={A small simulation and bootstrap workflow.},label={lst:simulation}]
library(regcorr)

set.seed(123)

res <- subRoutineTest(
  numSample = 100,
  p = 1,
  link = "1",
  model = "1",               # "1" bivariate normal, "2" Bernoulli
  betaTrue = c(0.2, 0.1),
  betaIni = c(0, 0),
  eta1True = c(0, 0),
  eta2True = c(0, 0),
  numSimu = 100,
  numBoot = 200
)

names(res)
res$RMSE
res$ConsistRate
res$power
\end{lstlisting}

The returned root mean squared error summarizes estimation accuracy, the consistency rate measures whether the estimated and true correlation effects agree in direction, and the empirical power summarizes how often the testing procedure detects the nonzero effect under the simulated design.

\section{Relationship to existing R tools}

The base R functions \texttt{cor()} and \texttt{cor.test()} estimate and test a fixed correlation coefficient. Packages for broader correlation analysis provide rich tools for robust, Bayesian, rank-based, partial, or multilevel correlation estimation. These tools answer questions about the magnitude or uncertainty of association. \texttt{regcorr} addresses a narrower but distinct question: how the Pearson correlation coefficient itself changes as a function of covariates.

This distinction matters. A partial correlation removes the linear effect of covariates from each response and then correlates residualized variables. In contrast, correlation regression directly specifies $\rho_i=h(x_i^\top\beta)$ and estimates how each covariate shifts the dependence structure. Thus, \texttt{regcorr} should be viewed as complementary to general correlation packages rather than as a replacement for them.

\section{Practical interpretation}

Interpreting coefficients in correlation regression requires attention to the link scale. Under the logistic link, $\rho_i$ is constrained to $(0,1)$, so the model is appropriate when the scientific assumption is that the two responses are positively associated. Under the hyperbolic tangent link, $\rho_i$ lies in $(-1,1)$ and can represent either positive or negative association. The sign of a coefficient describes movement on the link scale: for a monotone link, a positive coefficient increases the fitted correlation, and a negative coefficient decreases it.

It is also useful to report fitted correlations at representative covariate values. For example, rather than reporting only $\widehat{\beta}_1$, one may compute
\[
  \widehat{\rho}(x)=h(x^\top\widehat{\beta})
\]
for low, medium, and high covariate profiles. This turns the model output into a direct statement about the strength of association between the two responses.

\section{Availability}

The package is distributed on CRAN as \texttt{regcorr} and can be installed by
\begin{lstlisting}[style=rcode]
install.packages("regcorr")
\end{lstlisting}
The development repository is hosted at \url{https://github.com/lonze-nb/regcorr}. The CRAN release is version 0.1.0, licensed under MIT, and the package does not require compiled code.

\section{Discussion}

\texttt{regcorr} makes regression modeling of Pearson correlation coefficients available to R users through a compact set of fitting, simulation, and testing utilities. Its main use cases include simulation studies, methodological teaching, and applied analyses where the association between two responses is expected to vary across covariates. The package is especially suitable for researchers who want to move beyond a single global correlation estimate and examine heterogeneous association structures in a likelihood-based framework.

There are several natural extensions. First, future versions could provide formula interfaces that mirror familiar R modeling syntax, such as \texttt{y1 + y2 \textasciitilde{} x1 + x2}. Second, built-in summary, print, and plot methods could help users interpret fitted covariate-dependent correlations. Third, additional response types, such as mixed continuous-binary pairs or ordinal responses, would broaden the package's scope. Finally, simulation vignettes and applied case studies could further support reproducible usage.

By packaging a specialized statistical model into an accessible R implementation, \texttt{regcorr} provides a practical bridge between correlation-regression methodology and everyday statistical analysis.

\section*{Acknowledgements}

The authors thank the developers and maintainers of R and CRAN. The statistical framework implemented in \texttt{regcorr} is based on the work of Dufera, Liu, and Xu \cite{dufera2023}.

\section*{Conflict of interest}

The authors declare no conflict of interest.

\section*{Data and code availability}

The source code of \texttt{regcorr} is available from CRAN at \url{https://CRAN.R-project.org/package=regcorr} and from GitHub at \url{https://github.com/lonze-nb/regcorr}. The examples in this paper use simulated data generated by the package functions.

\end{document}